# Surface Transformations and Water Uptake on Liquid and Solid Butanol near the Melting Temperature

Panos Papagiannakopoulos[*,1,2], Xiangrui Kong[1], Erik S. Thomson[1], Nikola Markovic´[3], and Jan B. C. Pettersson[*,1]

[1] *Department of Chemistry and Molecular Biology, Atmospheric Science, University of Gothenburg, SE-412 96 Gothenburg, Sweden*

[2] *Department of Chemistry, Laboratory of Photochemistry and Kinetics, University of Crete, GR-71 003 Heraklion, Greece*

[3] *Department of Chemical and Biological Engineering, Physical Chemistry, Chalmers University of Technology, SE-412 96 Gothenburg, Sweden*

[*] To whom correspondence should be addressed. E-mail: panos@chem.gu.se (P.P.) or janp@chem.gu.se (J.B.C.P); Tel. +46 31 7869072; Fax: +46 31 7721394.

**Abstract**

Water interactions with organic surfaces are of central importance in biological systems and many Earth system processes. Here we describe experimental studies of water collisions and uptake kinetics on liquid and solid butanol from 160 to 200 K. Hyperthermal $D_2O$ molecules (0.32 eV) undergo efficient trapping on both solid and liquid butanol, and only a minor fraction scatters inelastically after an 80% loss of kinetic energy to surface modes. Trapped molecules either desorb within a few ms, or are taken up by the butanol phase during longer times. The water uptake and surface residence time increase with temperature above 180 K indicating melting of the butanol surface 4.5 K below the bulk melting temperature. Water uptake changes gradually across the melting point and trapped molecules are rapidly lost by diffusion into the liquid above 190 K. This indicates that liquid butanol maintains a surface phase with limited water permeability up to 5.5 K above the melting point. These surface observations are indicative of an incremental change from solid to liquid butanol over a range of 10 K straddling the bulk melting temperature, in contrast to the behavior of bulk butanol and previously studied materials.

# 1. Introduction

Water interactions with organic compounds are of central importance for biological systems[1] and many components of the Earth system[2] including organic aerosols in the atmosphere.[3] In biochemical processes, protein folding and function critically depend on the interactions with surrounding water molecules. Intermolecular interactions with water also result in clustering of hydrophobic molecules and contribute to the aggregation of lipids into membranes where hydrophilic groups interact with the surrounding water, while the hydrophobic termini remain hidden within the layer. The precise role of water in these important biological processes remains imperfectly understood, and resolving how water behaves near hydrophobic surfaces cannot rely on structure alone but must also consider the water dynamics.[1]

Water interactions also play a wide variety of roles in atmospheric processes.[2, 4] Primary emitted organic compounds are transformed in the atmosphere through a series of gas phase reactions, and low vapor pressure products condense on existing aerosol particles or are involved in the formation of new particles. Coating aerosols and cloud particles with organic trace species leads to modifications of their chemical and physical properties,[5-7] with potential impacts on climate vis-à-vis cloud formation and radiative forcing.[3, 4] In the atmosphere, primary hydrophobic compounds become increasingly hydrophilic through slow oxidation involving oxidizing agents like OH and ozone,[8] and thus the hygroscopicity of particles may change with age and oxidation state of the surface-bound compounds.[9] Secondary organic aerosol particles are usually assumed to be liquid, but they may also become solid or glassy with large resulting effects on gas uptake.[10]

Alcohols constitute one important group of atmospheric compounds and their competing hydrophobic and hydrophilic properties make them interesting as model systems for both atmospheric and biochemical systems. The smallest alcohol, methanol, is water soluble, even though detailed studies show that these small molecules tend to cluster in liquid water.[11] The rapid mixing of water and methanol is confirmed by molecular beam experiments where water is incorporated into a liquid methanol monolayer on the microsecond timescale already at temperatures below 200 K.[12] The solubility of the alcohols in water decreases rapidly with molecular size and although the four carbon molecule butanol still has a high solubility, octanol is only slightly soluble in water. Octanol-water partition coefficients are extensively used to predict processes of environmental and pharmacological importance, and the success of 1-octanol as a model solvent is generally attributed to its amphiphilic nature, which captures some of the complexity of real systems.[13] The bulk structure of 1-octanol consists of a diverse spectrum of aggregates, dominated by linear aggregates in dry octanol and by large micelle structures in water saturated octanol.[13, 14] In contrast to the random orientation in the bulk, molecular dynamics (MD) simulations show that octanol molecules near the surface become oriented into bilayers that at 298 K extend several molecular layers deep.[15] The water uptake and penetration into these types of surfaces is expected to depend sensitively on the bulk material properties and temperature, but the current understanding of the governing mechanisms is incomplete. Related MD simulations of organic systems suggest that penetration of water molecules into the interfacial region plays a significant role at water-oil interfaces,[16] and water permeation through model membranes can be significant on timescales of 100 ns and more.[17]

Alcohols are surface active on liquid and solid water and tend to form monolayers or multilayers that reduce water condensation and evaporation rates.[18] The effect depends on alcohol chain length and chain alcohols with 14 to 22 carbon atoms are reported to impede water evaporation by up to four orders of magnitude.[19, 20] Recent molecular beam experiments show that the effect of a methanol monolayer on ice is negligible, while a butanol layer on ice reduces uptake by 20-40% compared to pure ice.[21] Molecular dynamics simulations of butanol-covered liquid water suggest that water condensation is reduced by a factor of three at 300 K.[22] However, other studies on the water evaporation from supercooled sulfuric acid through butanol films show little to no effect of the butanol.[23, 24] These differing results suggest that the water interactions are quite sensitive to the detailed chemical and physical properties of adsorbed alcohol surface layers.

Detailed molecular level studies of system dynamics and kinetics often rely on spectroscopy and modeling, while the use of other sophisticated experimental methods including molecular beam techniques and surface science methods are usually hampered by the relatively high vapor pressures of the systems. The recent development of the environmental molecular beam (EMB) method makes studies of dynamics and kinetics possible at pressures up to $1\cdot10^{-2}$ mbar,[25, 26] and applications include studies of water interactions with thin alcohol layers on ice[21, 25] and graphite[12, 26]. Here we employ the EMB method for detailed studies of water interactions with solid and liquid n-butanol with the overall aim to characterize the mechanisms for water accommodation and bulk uptake. n-Butanol has a number of properties that makes it an interesting system for detailed studies. It is amphiphilic in nature and forms relatively strong hydrogen bonds. It is easily super-cooled but also has interesting solid phase properties at low temperature.[27-31] The EMB studies are carried out with $D_2O$ rather than $H_2O$ to enhance sensitivity in the experiments and cover the temperature from 160 to 200 K - thus including both solid and liquid bulk phases on either side of the 184.5 K bulk melting temperature of butanol.[32, 33] Of particular interest are the effects of surface structure on water uptake near the melting point. The implications for the understanding of water interactions with organic phases are discussed.

## 2. Experimental methods

### A. Experimental setup

The $D_2O$-butanol experiments were performed in an EMB apparatus; a six-chamber vacuum system that has been described in detail elsewhere.[25, 26] A molecular beam is generated by a pulsed gas source with part of the gas passing through a skimmer and a subsequent chopper to form a directed low density beam with square-wave-like 400 μs beam pulses. The beam source is run with a $D_2O$:He gas mixture at a total pressure of 2 bar and a partial $D_2O$ pressure of approximately 25 mbar, which produces a beam with mean kinetic energies of $0.32 \pm 0.02$ and $0.064 \pm 0.003$ eV for $D_2O$ and He, respectively.

The beam is directed towards a butanol-covered graphite surface (Advanced Ceramics Corp.; highly oriented pyrolytic graphite, grade ZYB, $12 \times 12$ mm$^2$) located in the center of the main ultra-high vacuum (UHV) chamber. The UHV chamber has a background pressure of approximately $10^{-9}$ mbar primarily due to residual background gases introduced during the experiments. In the EMB configuration the surface is surrounded by a separate inner environmental chamber that allows for experiments with vapor pressures into the $10^{-2}$ mbar range. The finite pressure distinguishes the method from traditional molecular beam experiments, and it has been termed EMB in analogy with environmental scanning electron microscopy. The apparatus has been designed to minimize the molecular beam path length (28 mm) within the high-pressure zone, such that the attenuation of the beam due to gas collisions within the inner chamber becomes significant only above $10^{-3}$ mbar.[25]

The incident $D_2O$/He beam enters the innermost chamber through a circular opening with a diameter of 5 mm and collides with the surface at an angle of 45°. The outgoing flux passes through a second 5 mm opening in the inner chamber wall and is monitored with a quadrupole mass spectrometer (QMS) at an angle of 45° from the surface normal direction. The QMS is rotatable and is also used to measure in the incident beam.

Micrometer thick *n*-butanol layers are produced on the graphite substrate by directly introducing butanol gas into the environmental chamber through a gas inlet. The thickness of the butanol layer is determined by monitoring the interference produced by the adsorbed layer when reflecting light from a diode laser (0.86 mW, 670 nm) at 3° from the surface normal direction.[34] A refractive index for *n*-butanol of 1.39577 at 670 nm is used to calculate the layer thickness.[35] Butanol layers are typically produced with an initial growth rate of approximately 70 monolayers per second (ML s$^{-1}$), and the butanol pressure is then adjusted to maintain a layer thickness of approximately 1 μm, corresponding to ca. 3000 ML (assuming that 1 ML of butanol consists of 3.69·10$^{14}$ molecules cm$^{-2}$ [36] and the unit cell volume of crystalline butanol is 3.735·10$^{-22}$ cm$^3$ [31]) The graphite surface is cleaned between experiments by heating to 500 K, and surface conditions are routinely confirmed by elastic helium scattering after surface cooling to 200 K or lower.[25, 26] Elastic helium scattering is also used to confirm that the graphite surface is completely covered with butanol during the experiments.[34]

**B. Analysis**

The time dependent flux from the surface measured by the QMS is recorded by a multi-channel scaler with a 10 μs dwell time, and the ion intensity counts are transformed into time-of-flight (TOF) distributions using the geometry of the system. The quantitative analysis of the TOF distributions relies on a nonlinear least-squares fitting of the measured intensities assuming the experimental data can be described by a combination of inelastic scattering (IS) and trapping followed by first-order thermal desorption (TD).[21] The thermal desorption is modeled with a residence time behavior of the form,

$$F_{res} = C_1 e^{-kt},$$
(1)

where $C_1$ is a scaling factor, $k$ is the desorption rate constant, and $t$ is time. The inelastic scattering distribution is assumed to have the common form,[37]

$$I_{IS}(v(t)) = C_2 v(t)^4 exp\left[-\left(\frac{v(t)-\bar{v}}{v_{IS}}\right)^2\right], \qquad (2)$$

where $C_2$ is a second scaling factor, $v(t)$ is the molecule velocity calculated from the travel time $t$ and flight path length $l$ between the surface and the QMS, $\bar{v}$ represents the peak of the inelastically scattered beam velocities, and $v_{IS}$ is,

$$v_{IS} = \sqrt{\frac{2k_B T_{IS}}{m}}, \qquad (3)$$

where the temperature $T_{IS}$ describes the velocity spread, $k_B$ is the Boltzmann constant, and $m$ is the molecular mass. The flexibility of the algorithm is ensured by the free fitting parameters $C_1$, $C_2$, $k$, $\bar{v}$ and $T_{IS}$.

The absolute probability for D$_2$O trapping followed by thermal desorption, $P_{TD}$, is computed by normalizing each thermal desorption integral by the thermal desorption integral from a contiguously measured bare graphite case. The trapping-desorption probability for hyperthermal D$_2$O scattering from graphite is constant within error limits in the temperature range used here, and the TD component has a cosine angular distribution independent of temperature.[38] The desorption for the bare case is then linearly scaled by the sticking coefficient $s_{graphite} = 0.73 \pm 0.07$ for D$_2$O on bare graphite under the present conditions.[21] Using this sticking coefficient as a scaling parameter $P_{TD}$ is easily computed as,

$$P_{TD} = s_{graphite} I_{TD}^{BuOH} / I_{TD}^{graphite}, \qquad (4)$$

where $I_{TD}^{BuOH}$ and $I_{TD}^{graphite}$ are the thermal desorption integrals in the butanol and graphite cases. In each case an error estimate is calculated from the range of integral values based on the 95% confidence intervals for the fitting parameters that contribute to the thermal decay function. The final error is based on propagating the error of each integral and the uncertainty in *s*.

## 3. Results

We have studied the dynamics and kinetics of $D_2O$ interactions with solid and liquid *n*-butanol from 160 to 200 K. The experimental results consist of TOF spectra that are further analyzed to determine desorption rate constants and the probabilities for inelastic scattering, thermal desorption and bulk uptake of water on butanol surfaces prepared by different procedures.

Figure 1 shows TOF distributions for $D_2O$ colliding with *n*-butanol at five different temperatures together with a distribution measured in the incident beam. A new butanol surface was prepared by vapor deposition before each experiment. Data points are overlaid by the two component non-linear least squares fitting of the TOF distributions with each component (IS and TD) and the sum of the two components is also depicted. The IS component is similar in all cases while the TD component changes rapidly with temperature. When increasing the temperature from 180 to 184 K, the TD distribution becomes wider indicating a longer residence time for adsorbed $D_2O$ on the butanol surface. The distribution continues to widen and the total TD intensity decreases as the temperature is further increased. Note that the melting temperature, $T_m$, for *n*-butanol is 184.5 K[32, 33], and thus the distributions at 184 and 186 K correspond to collisions with solid and liquid butanol, respectively. The TD distributions at 186 and 190 K are similar, but the TD component suddenly disappears at 191 K and higher temperatures.

The beam measurement displayed in Figure 1 is carried out with the same flight distance as the flux measurements from the surface, and the shape of the distribution therefore corresponds to hypothetical elastic scattering of $D_2O$ from the surface. The time shift between the incident beam distribution and the inelastic scattering peaks illustrates that water molecules loose kinetic energy during surface contact. The kinetic energy of $D_2O$ molecules directly scattered from the butanol surfaces was $0.064 \pm 0.006$ eV independent of temperature. This corresponds to 20% of the incident beam energy indicating substantial energy loss to surface modes in both liquid and solid butanol. Similar large energy losses have previously been observed for $D_2O$ collisions with butanol and methanol monolayers on water ice surfaces,[12, 21] as well as in gas collisions with water ice surfaces.[39, 40]

Figure 2 shows the absolute TD probability and the relative IS intensity as a function of temperature. The angular distribution for the TD component is described by a symmetric cosine distribution and measurements in a single direction are sufficient to calculate the total desorbing flux.[21] The absolute IS intensity can on the other hand not be determined since measurements are only carried out for a scattering angle of 45° and the angular distribution is unknown. The TD fraction varies widely with temperature, Figure 2a. The fraction increases from 0.4 at 160 K to 0.9 at 180 K, and then decreases rapidly above 180 K and levels out with values in the 0.2 – 0.35 range between 185 and 190 K. The TD fraction drops to very low values at 191 K and remains low up to 200 K. In sharp contrast to the TD results the IS intensity is more stable with relative values between 0.25 and 0.45, Figure 2b. The IS values begin to drop when temperature increases above 180 K and similar values are observed between 183 and 195 K, which indicates that the scattered $D_2O$ molecules interact with a similar surface both below and above the melting point. The observed TD probability of 0.9 at 180 K means that the IS intensity at the same $T$ cannot correspond to more than 10% of the total flux. The relatively small variation in IS over the entire temperature range indicates that the scattering probability remains low under all conditions. This conclusion is further supported by the large loss of kinetic energy during surface collisions (Figure 1), which is consistent with efficient trapping. We therefore conclude that the sharp decrease in TD intensity above 180 K (Figure 2a) is due to extensive loss of $D_2O$ to the butanol phase on the 10 ms timescale of the measurements. Isotopic exchange between $D_2O$ and surface $C_4H_9OH$ molecules was also considered as a possible sink of $D_2O$ molecules, but HDO desorption was experimentally confirmed to be below the detection level making this a less likely explanation.

The desorption rate constants determined from TOF data are presented as an Arrhenius diagram in Figure 3. The $k$ values clearly exhibit different behavior across the experimental temperature range. Starting with solid butanol at 180 K and lower, $k$ has an Arrhenius-type behavior and an activation energy $E_a = 0.08 \pm 0.03$ eV and a pre-exponential factor $A = 4.7 \cdot 10^{(5.0 \pm 0.8)}$ s$^{-1}$ are determined where error limits are 95% confidence intervals. The desorption rate constant rapidly decreases with temperature above 180 K, with the corresponding surface residence time $\tau = 1/k$ increasing from 250 μs at 180 K to about 500 μs at the bulk melting temperature. The desorption rate constant continues to decrease as the melting point is crossed, but now with a slightly different temperature dependence as illustrated in Figure 3. The apparent activation energies below and above the melting point are -0.57 ± 0.13 and -1.35 ± 0.08 eV, respectively. The surface residence time stabilizes at about 2 ms as the temperature approaches 190 K, and at higher $T$ water is rapidly lost by diffusion into the liquid butanol phase and no $k$ values could be determined. The combined results in Figures 1-3 clearly illustrate that water interactions with butanol changes significantly over a 10 K interval around the bulk melting point. The brief surface interactions experienced by inelastically scattered molecules are not strongly affected by melting (Figure 2b), while water molecules adsorbed on the μs to ms time scale are influenced by changes in surface properties (Figures 2a and 3). Thus, passing the bulk melting point does not have a singular effect, rather changes appear gradually between 180 and 190 K.

In the experiments described above, a new butanol layer was produced before each experiment started. To further investigate the effects of surface properties additional experiments were carried out where butanol layers were formed at specific temperatures and maintained as the temperature was changed in steps during cooling or heating. Figure 4 shows results from temperature ramps, where *i*) liquid butanol is deposited at 200 K and incrementally cooled to 160 K, and *ii*) solid butanol is deposited at 160 K and warmed to 190 K. Results from the "fresh" butanol studies described in Figures 1-3 are also included for comparison. Each measurement at an individual temperature took approximately 20 minutes with the intervening cooling and/or warming ramps proceeding at ca. 1 K min$^{-1}$. The desorption rate constants obtained during warming agree with the fresh butanol data, Figure 4a. This is also true when butanol is cooled from 200 K to the melting point, but at lower temperatures the $k$ values are always significantly lower for the cooled surface. This indicates that the surface properties of solid *n*-butanol, as expected, depend on the preparation procedure. However, below 180 K the calculated Arrhenius parameters for the warming ($E_a$ = 0.08 ± 0.05 eV, $A = 7.0 \cdot 10^{(5.0 \pm 1.3)}$ s$^{-1}$; $T$ = 160 – 175 K) and cooling ($E_a$ = 0.10 ± 0.03 eV, $A = 9.2 \cdot 10^{(5.0 \pm 0.8)}$ s$^{-1}$; $T$ = 160 – 180 K) cases agree within error limits with the data from freshly deposited butanol surfaces. Figure 4b shows that the TD values are similar for all three cases. The corresponding IS data are also comparable but not shown.

Earlier experiments have shown that *n*-butanol can be deeply super-cooled down to 125 K,[28, 30] and a super-cooled liquid layer could possibly form during the cooling experiments described above. The low desorption rate constants observed for the cooling case could indicate a longer residence time in a super-cooled liquid layer, but the TD values are on the other hand similar to the solid butanol case indicating limited uptake into the butanol phase. We performed independent low temperature tests where a liquid *n*-butanol layer was cooled to 170 K, and monitored for transient changes in water desorption behavior in excess of two hours. No time dependence was observed indicating that the surface conditions were stable on this time scale, and we have not conclusively determined if the surface is a super-cooled liquid or a polycrystalline solid under these conditions.

## 4. Discussion

The study clearly reveals that water collisions with both liquid and solid butanol are characterized by highly efficient energy transfer. Efficient energy transfer has previously been observed in similar systems including water interactions with thin liquid methanol films on graphite,[8] methanol and butanol monolayers on water ice,[21] and gases interacting with pure water ice surfaces.[39-40] For all of these systems, including solid and liquid butanol, a surface accommodation coefficient close to one can be expected under thermal conditions. On *n*-butanol, the trapped water molecules continue to either rapidly desorb, or become incorporated into the bulk during longer times. The competitive kinetics between the water desorption and uptake processes are obviously sensitive to the detailed surface conditions. In related work the competition between water desorption and bulk uptake was studied for methanol and butanol-covered water ice.[21] While a methanol layer does not constitute a barrier to water uptake on ice, a butanol monolayer does and a water uptake coefficient of about 0.8 was observed in the 155 – 200 K temperature range. The detailed mechanism for water uptake by alcohol-covered ice has not yet been identified, but it appears likely that thermal motion of butanol molecules will enhance the chances of water penetrating the surface layer. The alkyl chains of *n*-butanol prefer to orient upright on liquid water at 298 K, but butanol molecules are neither well-ordered nor evenly distributed making the surface layer porous, which enhances water penetration.[23, 41] Breakup of intermolecular butanol-butanol bonds also makes butanol OH groups available for hydrogen bonding and thereby enhances water stability on the surface. This is confirmed by MD simulations of butanol on liquid water at room temperature,[22] and is likely true also at lower temperatures. The same arguments apply to the neat butanol system studied here and we conclude that an increased water uptake and a decreased desorption probability are likely related to a larger mobility of butanol molecules in the surface layer.

The Arrhenius parameters observed for solid *n*-butanol below 180 K (e.g. $E_a = 0.08 \pm 0.03$ eV and $A = 4.7 \cdot 10^{(5.0 \pm 0.8)}$ s$^{-1}$ in Fig. 3) are low and unlikely to correspond to an ordinary desorption process. Considering the observed surface residence times in the μs – ms range and assuming desorption to be a first order process with a typical Arrhenius pre-exponential factor of $10^{13}$ s$^{-1}$, a surface binding energy of 0.3 – 0.4 eV would be expected. We have used MD simulation in order to estimate the water-butanol interaction energy employing the TIP4P/2005[42] potential for water and the OPLS-AA[43] potential for butanol. The butanol molecule was in its equilibrium geometry and both molecules were treated as rigid. Starting from random initial relative molecule-molecule geometries the potential energy was minimized using a simple velocity damping technique. Several local minima corresponding to water interaction with the hydrocarbon tail were found with well depths between 0.03 (water interacting with the C4 carbon) and 0.09 eV (water interacting with the C1 carbon), while hydrogen bonded structures had binding of energies of 0.25 or 0.34 eV. We also carried out a limited study of the interaction between a water molecule and a relaxed cluster of 11 butanol molecules. If the water molecule found hydroxyl groups to interact with the binding energy was 0.32 - 0.35 eV, otherwise 0.10 - 0.12 eV. Comparison with the experimental results suggests that the observed $E_a$ of 0.08 - 0.10 eV may correspond to the D$_2$O binding energy to an interface dominated by hydrophobic tails, while the low *A* values and associated long surface residence times indicate that adsorbed water molecules also form reversible hydrogen bonds with the surface. The resulting desorption kinetics are thus affected by transfer between at least two different types of bound states at the interface. This interpretation is consistent with our understanding of the solid butanol surface based on earlier Brillouin scattering, X-ray diffraction and calorimetric measurements.[29, 31] A state that consists of nanocrystallites embedded in a disordered matrix has been observed at 125 K - 160 K and a polycrystalline solid readily forms between 160 and 180 K. Considering the preparation procedure used here with rapid deposition of butanol we expect the solid butanol phase to be polycrystalline and the surface may be expected to be heterogeneous and relatively rough, which should make a range of surface sites with different properties available for water.

The most intriguing finding is undoubtedly the changes in water-butanol interactions observed between 180 and 190 K, which strongly suggest that the surface changes between solid and liquid over a 10 K interval around the melting point. Starting below the melting point, the water uptake and desorption kinetics begin to evolve around 180 K indicating changes in surface properties. One likely explanation for these changes is that surface melting occurs. That is, as the solid is warmed towards its melting temperature the surface becomes substantially disordered forming a thin liquid-like layer that grows thicker as the melting temperature is approached.[2, 44] This is a common phenomenon and numerous experimental and theoretical studies have demonstrated that water ice[44, 45] metals,[46] ceramics,[47] polymers[48] and colloids[49] undergo surface melting. Relatively few studies have been carried out with molecular solids and studies of short-chain organic compounds like *n*-butanol are apparently missing in the literature. Experimental studies of surface melting have been reported for solids consisting of small molecules including oxygen,[50] methane,[51] , biphenyl[52, 53] and caprolactam ($C_6H_{11}ON$),[54] and MD simulations of surface melting have been carried out for $SF_6$.[55] The results can usually be described by mean field theory which predicts that the film thickness increases as $(T_m - T)^{-1/3}$ for van der Waals solids, and as $\ln(T_m-T)$ for solids dominated by short range interactions.[45] An intriguing observation in the present study is that the apparent activation energy of -0.57 ± 0.13 eV observed for water desorption in the 180 – 184.5 K range (Figure 3) agrees fairly well with a $(T_m - T)^{-1/3}$ dependence near $T_m$, which suggests that the water residence time on *n*-butanol may scale with the thickness of the premelted surface layer. Previous work suggests polycrystallinity in the surface layer could contribute to a spread in the melting transition due to a combination of grain boundary energy and crystallite size effects.[45] However, here the desorption kinetics are well described by a single exponential decay of the $D_2O$ surface population. We therefore conclude that the whole surface is affected by the changes taking place since any heterogeneity in surface conditions would result in deviations from a single exponential decay. This makes us conclude that the results are not strongly affected by polycrystallinity in the sample.

The experimental results also suggest that although the butanol surface begins to liquefy below $T_m$, some molecular structure remains that limits water permeability up to 5.5 K above $T_m$. The effect disappears suddenly above 190 K, which indicates that it is due to a phase with specific properties rather than a preferential orientation of molecules at the surface that gradually disappears with increasing $T$. Long chain organic molecules such as alcohols[56] and alkanes[57] have previously been shown to form stable crystalline layers on top of the bulk liquid above the melting temperature, and the phenomenon has been termed surface freezing.[58] For alcohols $C_nH_mOH$ with chain lengths $16 < n < 28$ a crystalline surface bilayer is stable up to 1 K above $T_m$.[56] Water may intercalate into the center of the bilayer which results in bilayers stable up to 2 K above $T_m$ and observed for $n$ values down to 10.[56] These earlier studies show that alcohols have a tendency to form stable bilayers at the gas-liquid interface. Further evidence is provided by vibrational sum frequency generation spectroscopy (VSFG) which been used to investigate the organization of $n$-butanol at the air-liquid interface at 298 K.[59] VSFG spectra suggest that butanol molecules tend to aggregate into centrosymmetric structures on the liquid surface, and the data are consistent with butanol forming layered hydrogen-bonded structures at the interface. The effect is observed to be larger in hexanol than in butanol. Molecular dynamics simulations of the liquid $n$-octanol surface at 298 K showed that molecules tend to align perpendicular to the surface with the octyl groups pointing outwards.[15] The next layer preferentially orient antiparallel to the surface molecules resulting in a bilayer structure, and at least four hydrogen-bonded bilayers are identified beneath the surface before unsymmetric linear structures begin to dominate in the bulk.[13, 15] These previous studies suggest that the surface layer on liquid $n$-butanol may also have a bilayer structure. One suggestion is that the phase is a smectic liquid crystal phase,[36] but additional experimental and theoretical studies are required to test this hypothesis.

The current understanding of water uptake on atmospheric aerosol particles is far from complete. Although organic aerosol particles are often considered to be in a liquid state, it has recently been shown that organic aerosol particles can adopt an amorphous semisolid state that significantly influences gas exchange and heterogeneous reactivity.[60] The present study adds to this picture by suggesting that the surfaces of organics may have unique properties that further modify gas exchange in unpredictable ways. Is this behavior unique to small alcohols or even to $n$-butanol, or is it a common behavior also for other compounds? This provides the background for additional studies with more complex organic substances of environmental importance including alcohols, aldehydes and carboxylic acids.

## 5. Conclusions

We have studied $D_2O$ interactions with solid and liquid $n$-butanol in the temperature range from 160 to 200 K using EMB methods. The main conclusions can be summarized as follows:

- Hyperthermal collisions of $D_2O$ molecules with both solid and liquid butanol result in efficient trapping and only a minor fraction scatters inelastically after substantial loss (80%) of kinetic energy to surface modes.
- Inelastic scattering of $D_2O$ shows minor sensitivity to changes in surface properties in the 160 to 200 K range, including both solid and liquid butanol surfaces prepared under different conditions.
- A major fraction of the trapped water molecules either thermally desorbs within 0.25 – 3 ms, or is taken up by the butanol on a time scale longer than 10 ms. Desorption and uptake both depend on the temperature and surface properties of the butanol phase.
- Traversing the bulk melting point at 184.5 K does not have a dramatic effect on water uptake and desorption kinetics, and changes instead appear over a 10 K continuum between 180 and 190 K.
- Water uptake and surface residence time increase rapidly with increasing temperature above 180 K, indicating that solid butanol undergoes *surface melting* 4.5 K below the melting point.
- Liquid butanol maintains a surface layer with limited water permeability up to 5.5 K above the melting point. The permeability suddenly increases above 190 K and water is rapidly lost by diffusion into the bulk liquid phase. This indicates that the surface layer with limited water permeability corresponds to a distinct structure that is destroyed around 190 K.

The surface properties of butanol change gradually from solid to liquid over a 10 K interval centered around the bulk melting point, in sharp contrast to the behavior of the bulk material and with significant consequences for water uptake. Further studies should proceed to determine if the observed behavior of butanol is also found for other organic compounds, and theoretical studies including molecular dynamics simulations will help to further improve our understanding of water interactions with *n*-butanol and similar substances near melting. The EMB method has the potential of being further developed to allow for studies at higher pressures and temperatures[25] of benefit for the molecular level understanding of interface phenomena in biochemical and environmental systems.

**Acknowledgements**

This work is supported by the Swedish Research Council and the Nordic Top-Level Research Initiative CRAICC. PP thanks the Wenner-Gren Foundation for providing funding for an extended stay at the University of Gothenburg.

**Figure captions**

**Fig. 1** Time-of-flight distributions for $D_2O$ scattering and desorbing from liquid and solid *n*-butanol: experimental data (red points and line) and the total (black line), inelastic scattering (IS) (blue line) and thermal desorption (TD) (gray line) components of the non-linear fitting described in the text. The lower-most panel shows the distribution measured in the incident beam. The experimental data have been normalized to the incident beam intensity and smoothed with a seven point stepwise average. The surface temperature is indicated in each panel.

**Fig. 2** The absolute thermal desorption (TD) probability and the relative inelastic scattering (IS) intensity for $D_2O$ on *n*-butanol as a function of temperature. The dashed line indicates the melting temperature $T_m$ = 184.5 K for *n*-butanol.

**Fig. 3** Arrhenius plot of the desorption rate constant $k$ and the surface residence time $\tau$ for $D_2O$ on *n*-butanol. The solid line in the 160 – 180 K range is a linear least-square fit to the data with an activation energy $E_a$ = 0.08 ± 0.03 eV and a pre-exponential factor $A$ = 4.7·10$^{(5.0 \pm 0.8)}$ s$^{-1}$. Two additional linear least-square fits to the data are included in the 180 – 190 K range with apparent activation energies of -0.57 ± 0.13 and -1.35 ± 0.08 eV below and above the melting point, respectively. The dashed line indicates the melting temperature.

**Fig. 4** (a) The desorption rate constant $k$, and (b) the thermal desorption (TD) probability for $D_2O$ on *n*-butanol (BuOH) layers produced by different procedures: liquid butanol deposited at 200 K and cooled to 160 K (green), solid butanol deposited at 160 K and warmed to 190 K (purple), and new butanol layer deposited at each temperature (red) with data reproduced from Figures 1-3 without error limits. The dashed line indicates the melting temperature $T_m$ = 184.5 K for *n*-butanol.

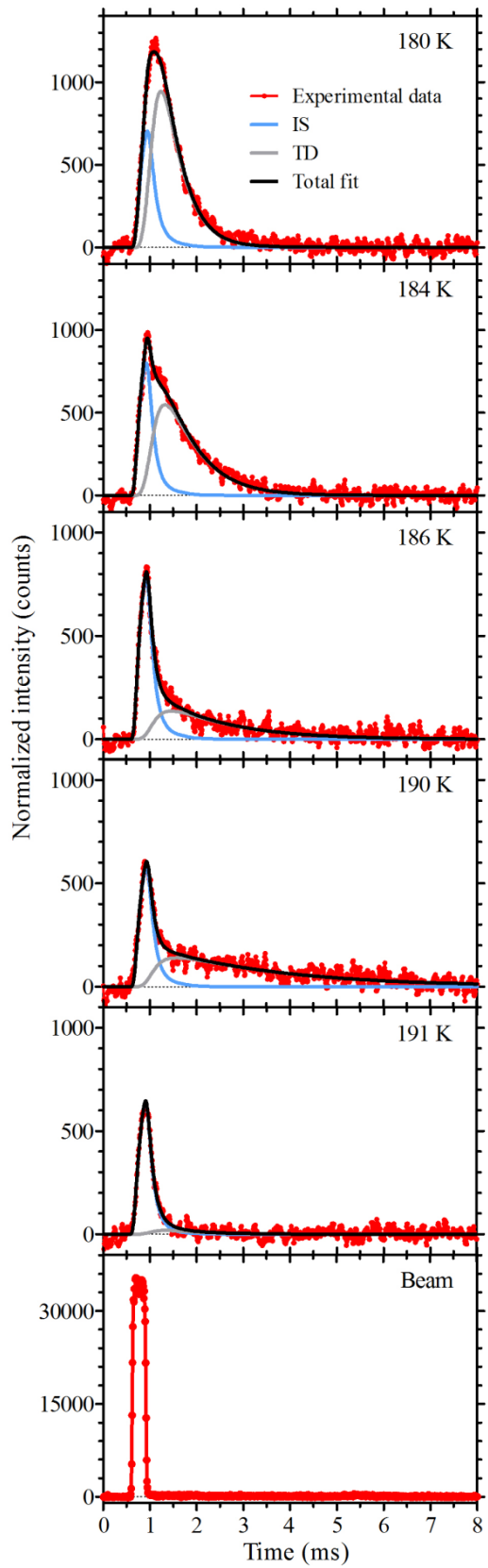

Figure 1

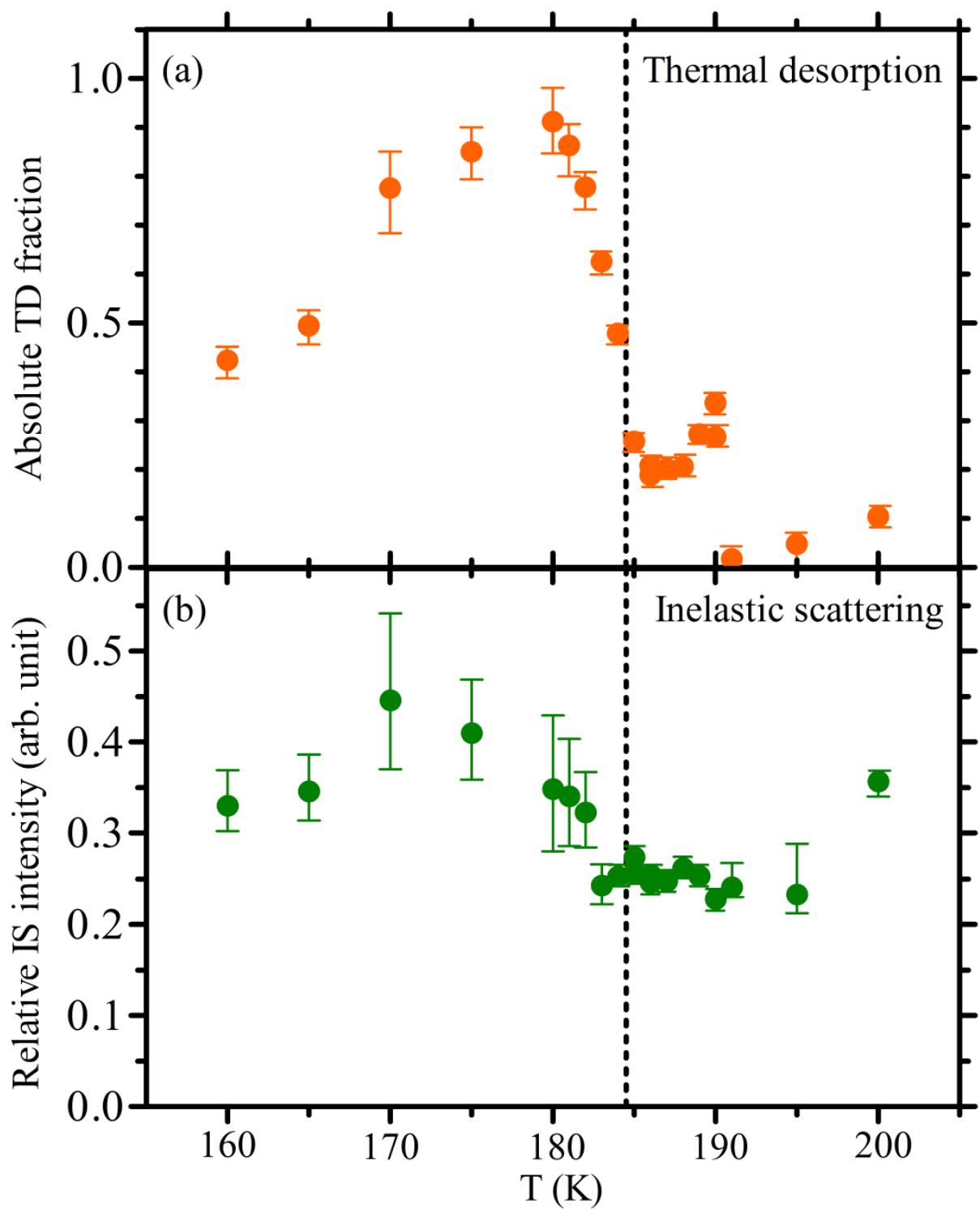

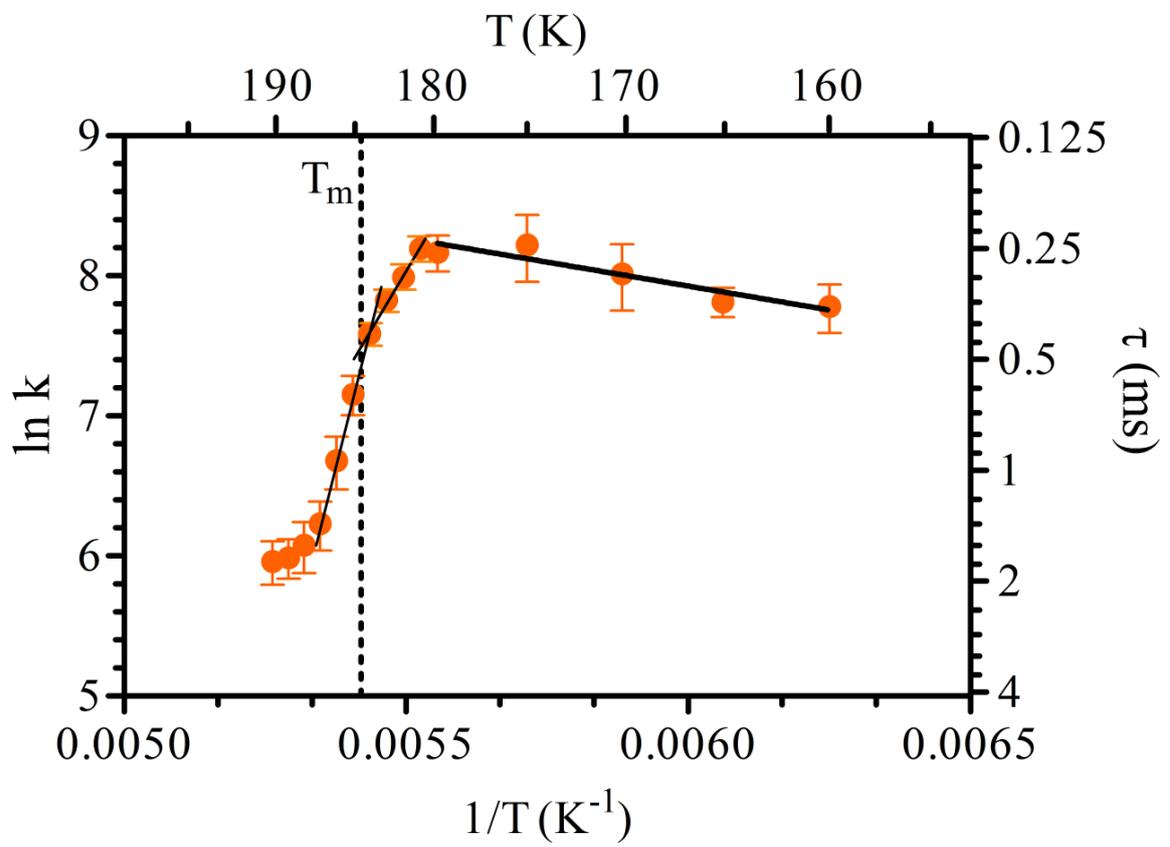

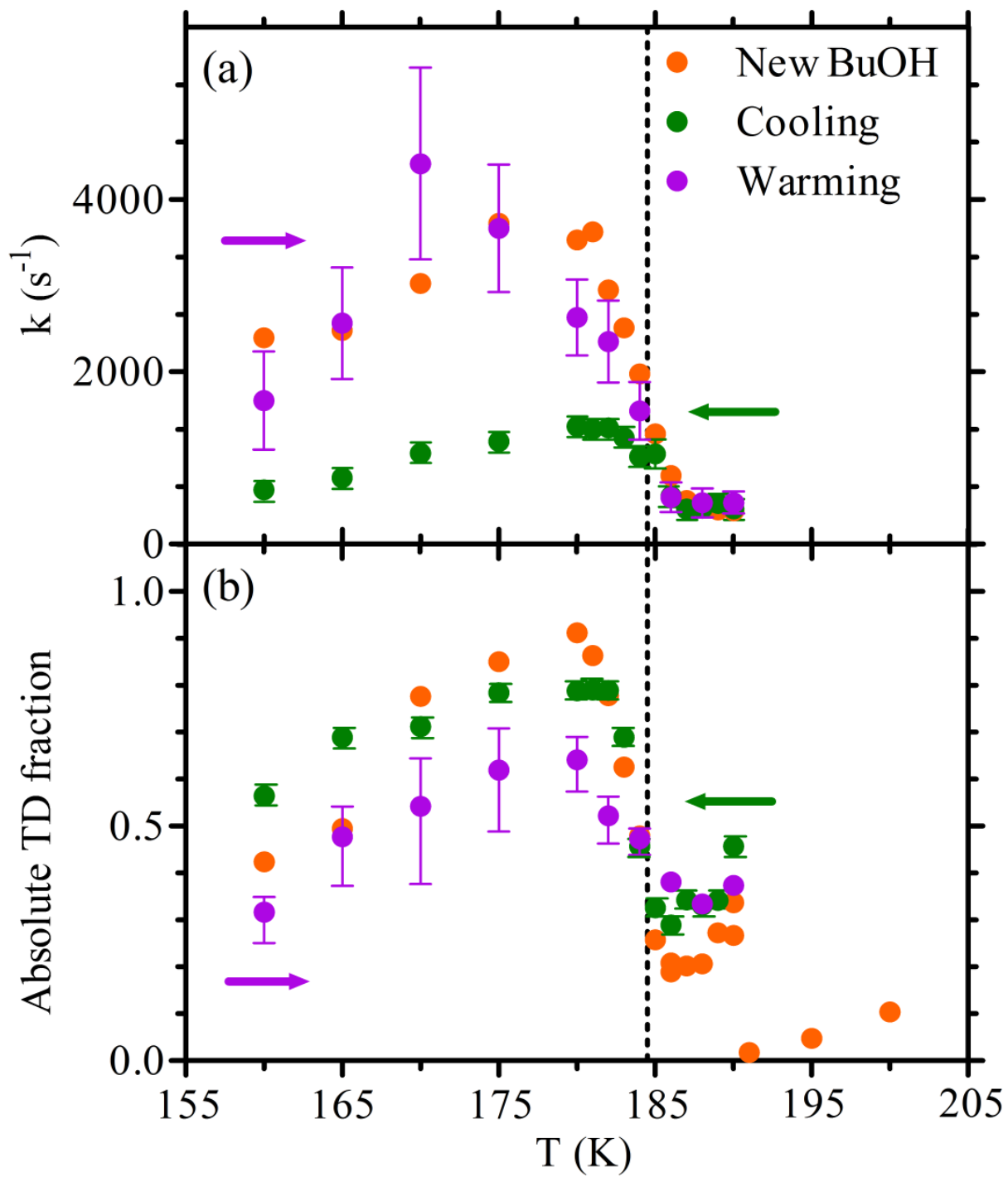

Figure 4

**Table of contents image**

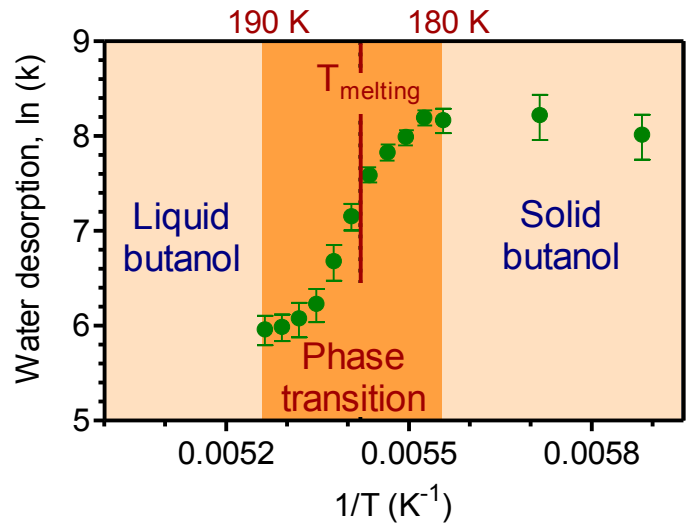